\documentclass{ws-procs9x6}
\newcommand{\bfk}{\mbox{\boldmath $k$}}
\newcommand{\bfp}{\mbox{\boldmath $p$}}
\newcommand{\pup}{p^\uparrow}
\newcommand{\be}{\begin{equation}}
\newcommand{\ee}{\end{equation}}
\newcommand{\bea}{\begin{eqnarray}}
\newcommand{\eea}{\end{eqnarray}}
\newcommand{\la}{\lambda}
\newcommand{\IoneG}{\mathcal{T}_{1}^{g}}
\newcommand{\ItwoG}{\mathcal{T}_{2}^{g}}
\def\nostrocostruttino#1\over#2{\mathrel{\mathop{\kern 0pt \rlap
{\hbox{$#1$}}} \hbox{\kern-.135em $#2$}}}
\def\sumint{\nostrocostruttino \sum \over {\displaystyle\int}}

\begin{document}

\title{Helicity Formalism and Spin Asymmetries in Hadronic Processes\footnote{$\,$
\uppercase{T}alk delivered by \uppercase{F}.~\uppercase{M}urgia at the
``\uppercase{I}nternational \uppercase{W}orkshop on \uppercase{T}ransverse
\uppercase{P}olarisation \uppercase{P}henomena in \uppercase{H}ard \uppercase{P}rocesses'',
\uppercase{TRANSVERSITY} 2005, \uppercase{S}eptember 7-10, 2005, \uppercase{C}omo,
\uppercase{I}taly. }}

\author{M. Anselmino,$^{\,\lowercase{a}}$ M. Boglione,$^{\,\lowercase{a}}$
U. D'Alesio,$^{\,\lowercase{b}}$ \\ E. Leader,$^{\,\lowercase{c}}$
 S. Melis,$^{\,\lowercase{b}}$ F. Murgia$^{\,\lowercase{b}}$}

\address{$^a$Dipartimento di Fisica Teorica, Universit\`{a} di Torino and\\
INFN, Sezione di Torino, V. P. Giuria 1, 10125 Torino, Italy\\
$^b$Dipartimento di Fisica, Universit\`{a} di Cagliari and\\ INFN, Sezione di Cagliari,
C.P. 170, 09042 Monserrato (CA), Italy\\
$^c$Imperial College London, Prince Consort Road, London SW7 2BW, U.K.} \maketitle
\abstracts{ We present a generalized QCD factorization scheme for the high energy inclusive
polarized process, $(A, S_A) + (B, S_B) \to C + X$, including all intrinsic partonic motions.
This introduces many non-trivial azimuthal phases and several new spin and $\bfk_\perp$
dependent soft functions. The formal expressions for single and double spin asymmetries are
discussed. Numerical results for $A_N(\pup \, p \to \pi \, X)$ are presented.}
\section{Introduction and formalism}\label{sec:intro}
Recently\cite{damu,abdlm,forma} we have developed an approach to study (un)polarized cross
sections for inclusive particle production in hadronic collisions at high energy and
moderately large $p_T$ and semi-inclusive deeply inelastic scattering (SIDIS).\cite{sidis}
Assuming that factorization is preserved, this approach generalizes the usual Leading Order
(LO), collinear perturbative QCD formalism by including spin and intrinsic transverse
momentum, $\bfk_\perp$, effects both in the soft contributions (parton distribution (PDF) and
fragmentation (FF) functions) and in the elementary processes. Helicity formalism is adopted
and exact non collinear kinematics is fully taken into account. Unpolarized cross sections
and transverse single spin asymmetries (SSA), with emphasis on the Sivers\cite{siv} and
Collins\cite{coll} effects, were already discussed in Refs.˜\ \refcite{damu,abdlm}. Here we
report on the most complete case of unpolarized cross sections and single and double spin
asymmetries for the process $(A, S_A) + (B, S_B) \to C + X$.\cite{forma} The cross section
for this process can be given as a LO (factorized) convolution of all possible hard
elementary QCD processes, $ab\to cd$, with soft, leading twist, spin and $\bfk_\perp$
dependent PDF and FF (see Eq.˜\ (8) of Ref. \refcite{abdlm}):
 \bea
 &&\frac{E_C \, d\sigma^{(A,S_A) + (B,S_B) \to C + X}} {d^{3} \bfp_C} =\nonumber \\
 && \sum_{a,b,c,d, \{\la\}}\,\,\,
  \int \frac{dx_a \, dx_b \, dz}{16 \pi^2 x_a x_b z^2  s} \;
 d^2 \bfk_{\perp a} \, d^2 \bfk_{\perp b}\, d^3 \bfk_{\perp C}\, \delta(\bm{k}_{\perp C} \cdot
 \hat{\bm{p}}_c) \, J(\bm{k}_{\perp C}) \nonumber \\
 &&\qquad\qquad\times \,\, \rho_{\la^{\,}_a, \la^{\prime}_a}^{a/A,S_A} \, \hat f_{a/A,S_A}(x_a,\bfk_{\perp a})
 \> \rho_{\la^{\,}_b, \la^{\prime}_b}^{b/B,S_B} \,
 \hat f_{b/B,S_B}(x_b,\bfk_{\perp b}) \label{gen1} \\
 &&\qquad\qquad\times \,\, \hat M_{\la^{\,}_c, \la^{\,}_d; \la^{\,}_a, \la^{\,}_b} \, \hat M^*_{\la^{\prime}_c,
 \la^{\,}_d; \la^{\prime}_a, \la^{\prime}_b} \> \delta(\hat s + \hat t + \hat u) \> \hat
 D^{\la^{\,}_C,\la^{\,}_C}_{\la^{\,}_c,\la^{\prime}_c}(z,\bfk_{\perp C}) \>, \nonumber
\eea
 \noindent where $A$, $B$ are initial, spin 1/2 hadrons in pure spin states $S_A$ and
$S_B$; $C$ is the unpolarized observed hadron; $J(\bfk_{\perp C})$ is a phase-space
kinematical factor;\cite{damu} $\rho_{\la^{\,}_a, \la^{\prime}_a}^{a/A,S_A} \, \hat
f_{a/A,S_A}(x_a,\bfk_{\perp a})$ contains all information on parton $a$ and its polarization
state, through its helicity density matrix and the spin and $\bfk_\perp$ dependent PDF
(analogously for parton $b$); $\hat M_{\la^{\,}_c, \la^{\,}_d; \la^{\,}_a, \la^{\,}_b}$ are
the LO helicity amplitudes for the elementary process $ab\to cd$;
$D^{\la^{\,}_C,\la^{\,}_C}_{\la^{\,}_c,\la^{\prime}_c}(z,\bfk_{\perp C})$ is a product of
soft helicity fragmentation amplitudes for the $c\to C+X$ process. The remaining notation, in
particular for kinematical variables, should be obvious.\cite{damu,abdlm} Let us stress here
that a formal proof of factorization for the $AB\to C+X$ process in the non collinear case is
still missing; universality and evolution properties of the new spin and $\bfk_\perp$
dependent PDF and FF are not established or well known yet; a consistent account of all
higher-twist effects is still missing. In the sequel, we will discuss in more detail the
basic ingredients of Eq.˜\ (\ref{gen1}).
\section{Spin and $\bfk_\perp$ dependent PDF and FF (leading twist)}\label{sec:pdff}
 The most general expression for the helicity density matrix of quark $a$ inside hadron $A$ with
polarization state $S_A$ is
 \be
 \rho_{\la^{\,}_a, \la^{\prime}_a}^{a/A,S_A}  = \frac{1}{2}\,{\left(
 \begin{array}{cc}
 1+P^a_z & P^a_x - i P^a_y \\
 P^a_x + i P^a_y & 1-P^a_z
 \end{array}
 \right)}_{\!\!\!\!A,S_A} \!\!\!\!\!\!\! = \frac{1}{2}\,{\left(
 \begin{array}{cc}
 1+P^a_L & P^a_T \, e^{-i\phi_{s_a}} \\
 P^a_T \, e^{i\phi_{s_a}} & 1-P^a_L
 \end{array}
 \right)}_{\!\!\!\!A,S_A}\!\!\! , \label{rho-a}
 \ee
where $P^a_j$ is the $j$-th component of the quark polarization vector in its helicity frame.
Introducing soft, nonperturbative helicity amplitudes for the inclusive process $A\to a+X$,
${\hat{\mathcal F}}_{\la^{\,}_a, \la^{\,}_{X_A}; \la^{\,}_A}$, we can write
 \bea
 \rho_{\la^{\,}_a, \la^{\prime}_a}^{a/A,S_A} \> \hat f_{a/A,S_A}(x_a,\bfk_{\perp a}) &=&
 \sum _{\la^{\,}_A, \la^{\prime}_A} \rho_{\la^{\,}_A, \la^{\prime}_A}^{A,S_A} \sumint_{X_A,
 \la_{X_A}} \!\!\!\!\! {\hat{\mathcal F}}_{\la^{\,}_a, \la^{\,}_{X_A}; \la^{\,}_A} \, {\hat{\mathcal
 F}}^*_{\la^{\prime}_a,\la^{\,}_{X_A}; \la^{\prime}_A} \nonumber \\
 &\equiv& \sum _{\la^{\,}_A, \la^{\prime}_A} \rho_{\la^{\,}_A, \la^{\prime}_A}^{A,S_A} \> \hat
 F_{\la^{\,}_A, \la^{\prime}_A}^{\la^{\,}_a,\la^{\prime}_a} \>,\label{defF} \eea
where $\rho_{\la^{\,}_A, \la^{\prime}_A}^{A,S_A}$ is in turn the helicity density matrix of
hadron $A$
 \be
  \rho_{\la^{\,}_A, \la^{\prime}_A}^{A,S_A} = \frac{1}{2}\,{\left(
\begin{array}{cc}
1+P^A_Z & P^A_X - i P^A_Y \\
 P^A_X + i P^A_Y & 1-P^A_Z
\end{array}
\right)}= \frac{1}{2}\,{\left(
\begin{array}{cc}
1+P^A_L & P^A_T \, e^{-i\phi_{S_A}}  \\
P^A_T \, e^{i\phi_{S_A}} & 1-P^A_L
\end{array}
\right)} \,, \label{rho-A}
 \ee
 and $P^A_J$ is the $J$-th component of the
hadron polarization vector in its helicity rest frame. The definition of $\hat F_{\la^{\,}_A,
\la^{\prime}_A}^{\la^{\,}_a,\la^{\prime}_a}$ in terms of the helicity amplitudes
${\hat{\mathcal F}}_{\la^{\,}_a, \la^{\,}_{X_A}; \la^{\,}_A}$ can be deduced from Eq.˜\
(\ref{defF}). One can see that, due to rotational invariance and parity properties, the
following relations hold:\cite{lead}
 \bea
 \hat F_{\la^{\prime}_A, \la^{\,}_A}^{\la^{\prime}_a,\la^{\,}_a} &=&
 \left(\,\hat F_{\la^{\,}_A, \la^{\prime}_A}^{\la^{\,}_a,\la^{\prime}_a}\,\right)^*\,,
 \nonumber\\
 \hat F_{\la^{\,}_A, \la^{\prime}_A}^{\la^{\,}_a,\la^{\prime}_a}(x_a,\bfk_{\perp a})
 &=& F_{\la^{\,}_A, \la^{\prime}_A}^{\la^{\,}_a,\la^{\prime}_a}(x_a,k_{\perp a})\,
 \exp\left[\,i(\la^{\,}_A-\la^{\prime}_A)\phi_a\,\right]\,, \label{ffprop}\\
 {F}_{-\la^{\,}_{A},-\la^{\prime}_A}^{ -\la^{\,}_a,-\la^{\prime}_{a}}
 &=& (-1)^{2(S_A -s_a)} \> (-1)^{(\la^{\,}_A -\la^{\,}_a) + (\la^{\prime}_A -\la^{\prime}_a)}
\> {F}_{\la^{\,}_{A},\la^{\prime}_A}^{ \la^{\,}_a,\la^{\prime}_{a}} \,. \nonumber\eea
 Using Eq.˜\ (\ref{ffprop}) one can easily associate the eight functions \be
 F^{++}_{++},\quad F^{++}_{--},\quad F^{+-}_{+-},\quad F^{+-}_{-+},\quad
 F^{++}_{+-},\quad F^{--}_{+-},\quad F^{+-}_{++},\quad F^{+-}_{--}
 \,,\label{ff8}
\ee to the eight leading twist, spin and $\bfk_\perp$ dependent PDF: $F^{++}_{++} \pm
F^{++}_{--}$ are respectively related to the unpolarized and longitudinally polarized PDF,
$f_{a/A}$ and $\Delta_L f_{a/A}$; $F^{+-}_{+-}\pm F^{+-}_{-+}$ are related with the two
possible contributions to the transversity PDF; $F^{++}_{+-} \pm F^{--}_{+-}$ are
respectively related to the probability of finding an unpolarized (longitudinally polarized)
parton inside a transversely polarized hadron, the Sivers function,\cite{siv}
$\Delta^{\!N}f_{a/A^\uparrow}$ (the $g^\perp_{1T}$ PDF); $F^{+-}_{++}\pm F^{+-}_{--}$ are
related to the probability of finding a transversely polarized parton respectively inside an
unpolarized hadron (the so-called Boer-Mulders function,\cite{boemu}
$\Delta^{\!N}f_{a^\uparrow/A}$) and inside a longitudinally polarized hadron, the
$h^\perp_{1L}$ PDF. More precisely, the relations are the following:\cite{forma}
 \bea
 \hat f_{a/A} &=& \hat f_{a/A,S_L} = \left( F^{++}_{++} + F^{++}_{--} \right)\nonumber \\
 \hat f_{a/A,S_T} &=& \hat f_{a/A} + \frac{1}{2}\,\Delta\hat f_{a/S_T} =
 \left( F^{++}_{++} + F^{++}_{--} \right) + 2 \, {\rm Im} F^{++}_{+-}
 \sin(\phi_{S_A} -\phi_a)\nonumber\\
 P_x^a \, \hat f_{a/A,S_L} &=& \Delta\hat f_{s_x/S_L} = 2 \, {\rm Re} F^{+-}_{++}\nonumber\\
 P_x^a \, \hat f_{a/A,S_T} &=& \Delta\hat f_{s_x/S_T} = \left( F^{+-}_{+-} + F^{-+}_{+-} \right) \,
 \cos(\phi_{S_A}-\phi_a)
 \label{A23}\\
 P_y^a \, \hat f_{a/A,S_L} &=& P_y^a \, \hat f_{a/A} = \Delta\hat f_{s_y/S_L}
 = -2 \, {\rm Im}  F^{+-}_{++} \nonumber\\
 P_y^a \, \hat f_{a/A,S_T} &=& \Delta\hat f_{s_y/S_T} =
 -2 \, {\rm Im}  F^{+-}_{++} + \left( F^{+-}_{+-} - F^{-+}_{+-}
 \right) \, \sin(\phi_{S_A}-\phi_a)
 \nonumber \\
 P_z^a \,\hat f _{a/A,S_L} &=& \Delta\hat f_{s_z/S_L} =
 \left( F^{++}_{++} - F^{++}_{--} \right) \nonumber\\
 P_z^a \,\hat f _{a/A,S_T} &=& \Delta\hat f_{s_z/S_T} =
 2 \, {\rm Re} F^{++}_{+-} \cos(\phi_{S_A} -\phi_a) \,, \nonumber
 \eea
where $\phi_{S_A}$ and $\phi_a$ are respectively the azimuthal angle of the hadron spin
polarization vector and of the parton $a$ transverse momentum, $\bfk_{\perp a}$, in the
hadronic c.m. frame. We have also used the notation $\Delta\hat f_{s_i/S_J}=\hat
f_{s_i/S_J}-\hat f_{-s_i/S_J}$. More details and relations with the notation of the Amsterdam
group\cite{boemu} can be found in Ref.˜\ \refcite{forma}. Since the helicity density matrix
for a massless gluon can be formally written similarly to that of a quark,
 \be
 \rho_{\lambda_g^{\,},
\lambda^{\prime}_g}^{g/A,S_A}= \frac{1}{2}{\left(
\begin{array}{cc}
1+P_{z}^{g} &
\IoneG-i\ItwoG \\
\IoneG+i \ItwoG & 1-P_{z}^{g}
\end{array}
\right)}_{\!\!\!\!A,S_A} \!\!\!\!\!\!\!\! = \frac{1}{2}{\left(
\begin{array}{cc}
1+ P^g_{circ}&
- P^g_{lin} \, e^{-2i\phi}\\
- P^g_{lin} \, e^{2i\phi} & 1-P^g_{circ}
\end{array}
\right)}_{\!\!\!\!A,S_A}\!\! \label{rho-gl} \!\!\!\!\!\!, \ee where $\IoneG$ and $\ItwoG$ are
related to the degree of linear polarization of the gluon, relations analogous to those shown
for quarks hold also for gluons.\cite{forma}

 Analogously, introducing soft nonperturbative
helicity fragmentation amplitudes for the process $c\to C+X$, and limiting to the case of
unpolarized hadron $C$, properties similar to those shown for PDF in Eq.˜\ (\ref{ffprop})
hold.\cite{abdlm,forma}
 {}From these relations one can easily see that, for each parton, only two independent FF survive:
 the usual unpolarized FF; the well known Collins function\cite{coll} for quarks,
 $\Delta^{\!N}\hat{D}_{C/q^\uparrow}(z,k_{\perp C})$, and a Collins-like function for
(linearly) polarized gluons, $\Delta^{\!N}\hat{D}_{C/\IoneG}(z,k_{\perp C})$.\cite{forma}
\section{Helicity amplitudes for the elementary process $ab\to cd$}\label{sec:he}
Since intrinsic parton motions are fully taken into account in our approach, all soft
processes, $A(B) \to a(b)+X$ and $c\to C+X$, and the elementary process $ab\to cd$ take place
out of the hadronic production plane (the $XZ_{cm}$ plane). The relation between the
elementary helicity amplitudes given in the hadronic c.m. frame, $\hat M_{\la^{\,}_c,
\la^{\,}_d; \la^{\,}_a, \la^{\,}_b}$, and those given in the canonical partonic c.m. frame
(no azimuthal phases), $\hat M^0_{\la^{\,}_c, \la^{\,}_d; \la^{\,}_a, \la^{\,}_b}$, has been
given in Ref.˜\ \refcite{abdlm}. In summary,
 \bea
 \hat M_{\la^{\,}_c, \la^{\,}_d; \la^{\,}_a, \la^{\,}_b} \! &=& \hat M^0 _{\la^{\,}_c,
 \la^{\,}_d; \la^{\,}_a, \la^{\,}_b} \label{M-M0} \\
 &\times& \, e^{-i (\la^{\,}_a \xi _a + \la^{\,}_b \xi _b -
          \la^{\,}_c \xi _c - \la^{\,}_d \xi _d)}
 \, e^{-i [(\la^{\,}_a - \la^{\,}_b) \tilde \xi _a -
         (\la^{\,}_c - \la^{\,}_d) \tilde \xi _c]}
 \, e^{i(\la^{\,}_a - \la^{\,}_b)\phi^{\prime\prime}_c}\,, \nonumber
\eea where $\xi_j$, $\tilde{\xi}_j$ ($j=a,b,c,d$), $\phi_c''$ are phases which depend on the
initial kinematical configuration in the hadronic c.m. frame.\cite{abdlm} Parity properties
for the canonical helicity amplitudes $\hat M^0$ are well known,\cite{lead} and so are the
relations between a given canonical helicity amplitude and those obtained by exchanging the
two initial (final) partons.\cite{lead} For massless partons there are only three independent
helicity amplitudes, $\hat{M}_{++;++}=\hat{M}^0_1\,\exp(i\varphi_1)$,
$\hat{M}_{-+;-+}=\hat{M}^0_2\,\exp(i\varphi_2)$,
$\hat{M}_{-+;+-}=\hat{M}^0_3\,\exp(i\varphi_3)$, where $\varphi_j$ ($j=1,2,3)$ are the
corresponding phases given in Eq. (\ref{M-M0}). At LO there are eight elementary
contributions $ab\to cd$ which must be considered separately, since they involve different
combinations of PDF and FF in Eq.˜\ (\ref{gen1}): $q_a q_b \to q_c q_d$, $g_a g_b \to g_c
g_d$, $q g \to q g$, $g q \to g q$, $q g \to g q$, $g q \to q g$, $g_a g_b \to q \bar{q}$, $q
\bar{q} \to g_c g_d$ (the first contribution includes all quark and antiquark cases).
\section{Kernels for the process $A(S_A)+B(S_B)\to C+X$}\label{sec:kern}
In the previous sections we have presented all the ingredients required for the evaluation of
the convolution integral for the double polarized cross section, Eq.˜\ (\ref{gen1}). We can
then derive the expression of all the physically relevant single and double spin asymmetries
for the process $A(S_A)+B(S_B)\to C+X$, and, with appropriate modifications, for other
inclusive production processes. Defining kernels as:
 \bea
 \Sigma(S_A,S_B)^{ab\to cd} &=& \sum _{\{\lambda\}} \rho_{\la^{\,}_a,
 \la^{\prime}_a}^{a/A,S_A} \, \hat f_{a/A,S_A}(x_a,\bfk_{\perp a}) \> \rho_{\la^{\,}_b,
 \la^{\prime}_b}^{b/B,S_B} \, \hat f_{b/B,S_B}(x_b,\bfk_{\perp b})
 \nonumber\\
 &\times& \hat M_{\la^{\,}_c, \la^{\,}_d; \la^{\,}_a, \la^{\,}_b} \, \hat M^*_{\la^{\prime}_c,
 \la^{\,}_d; \la^{\prime}_a, \la^{\prime}_b} \> \hat
 D^{\la^{\,}_C,\la^{\,}_C}_{\la^{\,}_c,\la^{\prime}_c}(z,\bfk_{\perp C})\,,\label{kern}
\eea
 we present here, as an example, the kernel for the
$q_aq_b\to q_cq_d$ process:\bea
 && \Sigma(S_A,S_B)^{q_a q_b \to q_c q_d} = \frac 12 \, \hat D_{C/c}(z, k_{\perp C}) \>
 \hat f_{a/S_A}(x_a, \bfk_{\perp a}) \> \hat f_{b/S_B}(x_b, \bfk_{\perp b}) \nonumber \\
 &\times& \Biggl\{ \left( |\hat M^0_1|^2 + |\hat M^0_2|^2 + |\hat M^0_3|^2
 \right) + P_z^a \, P_z^b \left( |\hat M^0_1|^2 - |\hat M^0_2|^2 - |\hat M^0_3|^2
 \right) \nonumber \\
 &+& \hat M^0_2 \, \hat M^0_3 \left[ \left( P_x^a P_x^b +
 P_y^a P_y^b \right) \, \cos(\varphi_{3}-\varphi_{2}) - \left( P_x^a P_y^b - P_y^a P_x^b
 \right) \, \sin(\varphi_{3}-\varphi_{2})
 \right] \Biggr\} \nonumber \\
 &-& \frac 12 \, \Delta^N\hat D_{C/c^\uparrow}(z, k_{\perp C}) \>
 \hat f_{a/S_A}(x_a, \bfk_{\perp a}) \> \hat f_{b/S_B}(x_b, \bfk_{\perp b}) \label{qqqq} \\
 &\times& \Biggl\{ \hat M^0_1 \, \hat M^0_2 \, \left[
   P_x^a \, \sin(\varphi_1 - \varphi_2 + \phi_C^H)
 - P_y^a \, \cos(\varphi_1 - \varphi_2 + \phi_C^H) \right] \nonumber \\
 &+& \hat M^0_1 \, \hat M^0_3 \, \left[
   P_x^b \, \sin(\varphi_1 - \varphi_3 + \phi_C^H)
 - P_y^b \, \cos(\varphi_1 - \varphi_3 + \phi_C^H) \right] \Biggr\} \>, \nonumber
\eea where $\phi_C^H$ is the azimuthal angle of hadron $C$ three-momentum in the helicity
frame of parton $c$. A more complete list of kernels for all partonic contributions and more
details can be found in Ref.˜\ \refcite{forma}.
\section{Cross section and SSA for the process $pp\to\pi+X$}\label{sec:ssa}
The formalism described in the previous sections is very general and can be applied to the
calculation of unpolarized cross sections, single and double spin asymmetries for inclusive
particle production in hadronic collisions. As an explicit example, we now discuss the
transverse single spin asymmetry for inclusive pion production in proton-proton collisions,
$A_N(pp\to\pi+X)=(d\sigma^\uparrow-d\sigma^\downarrow)/(d\sigma^\uparrow+d\sigma^\downarrow)$,
where $d\sigma^{\uparrow,\downarrow}$ stands for the cross section of Eq. (\ref{gen1}) with
$S_A=\uparrow,\downarrow$, $S_B=0$. We present, limiting ourselves to the case $q_aq_b\to
q_cq_d$, the combination of kernels appearing in the numerator and denominator of the SSA
(dependence on $x_{a,b}$, $\bfk_{\perp a,b}$ in PDF and on $z$, $\bfk_{\perp C}$ in FF is
understood):
 \bea
 &&[\Sigma(\uparrow,0) - \Sigma(\downarrow,0)]^{q_a q_b \to q_c q_d} = \nonumber \\
 &\,& \frac{1}{2} \, \Delta \hat f_{a/A^\uparrow} \, \hat f_{b/B}
 \, \left[\,|{\hat M}_1^0|^2 + |{\hat M}_2^0|^2 + |{\hat M}_3^0|^2 \right] \,
 \hat D _{C/c} \nonumber \\
 &&+\,  2\,\left[ \Delta^- \hat f^a_{s_y/\uparrow} \, \cos(\varphi_3 -\varphi_2)
 -\Delta \hat f^a_{s_x/\uparrow}\, \sin(\varphi_3 -\varphi_2) \right]
 \, \Delta \hat f^b_{s_y/B}\, {\hat M}_2^0 \, {\hat M}_3^0 \,\hat D_{C/c}
 \nonumber \\
 &&+\, \left[ \Delta^- \hat f^a_{s_y/\uparrow}\, \cos(\varphi_1 -\varphi_2 +
 \phi_C^H) - \Delta \hat f^a_{s_x/\uparrow}\,
 \sin(\varphi_1 -\varphi_2 + \phi_C^H) \right] \,\label{num-asym-qq} \\
 &&\times \,\, \hat f_{b/B}\, {\hat M}_1^0 \, {\hat M}_2^0 \, \Delta^N {\hat
 D}_{C/c^\uparrow}\,
 \nonumber \\
 &&+\,  \frac{1}{2} \, \Delta \hat f_{a/A^\uparrow}\, \Delta \hat f^b_{s_y/B} \,
  \cos(\varphi_1 -\varphi_3 + \phi_C^H) \, {\hat M}_1^0 \, {\hat M}_3^0 \,
 \Delta ^N {\hat D}_{C/c^\uparrow} \nonumber\,,
\eea
 \bea
 &&[\Sigma(\uparrow,0) + \Sigma(\downarrow,0)]^{q_aq_b\to q_cq_d} =
 \nonumber \\
 &\,& \hat f_{a/A} \, \hat f_{b/B}  \, \left[ \, |{\hat
 M}_1^0|^2 + |{\hat M}_2^0|^2 + |{\hat M}_3^0|^2 \right] \,
 \hat D _{C/c}\nonumber \\
 &&+\,  2\, \Delta \hat f^a_{s_y/A} \,
         \Delta \hat f^b_{s_y/B} \,
 \cos(\varphi_3 -\varphi_2)\, {\hat M}_2^0 \, {\hat M}_3^0 \,\hat D _{C/c}
 \label{den-asym-qq} \\
 &&+\, \left[ \hat f_{a/A}\, \Delta \hat f^b_{s_y/B}\,
 \cos(\varphi_1 -\varphi_3 + \phi_C^H)\,
 {\hat M}_1^0 \, {\hat M}_3^0 \nonumber \right. \\
 && \! + \, \left. \Delta \hat f^a_{s_y/A}\, \hat f_{b/B}\,
 \cos(\varphi_1 -\varphi_2 + \phi_C^H)\, {\hat M}_1^0 {\hat M}_2^0\, \right] \, \Delta ^N
 {\hat D} _{C/c^\uparrow}\nonumber\,.
\eea There are four terms contributing to the numerator of the SSA, Eq.˜(\ref{num-asym-qq}):
the Sivers contribution (2nd line); the transversity$\otimes$Boer-Mulders contribution (3rd
line); the transversity$\otimes$Collins contribution (4th and 5th lines); the
Sivers$\otimes$Boer-Mulders$\otimes$Collins contribution (last line). Similarly, there are
three terms contributing to the denominator of the SSA, Eq.˜(\ref{den-asym-qq}): the usual
term involving only unpolarized quantities (2nd line); the Boer-Mulders$\otimes$Boer-Mulders
contribution (3rd line); the Boer-Mulders$\otimes$Collins contribution (last two lines).
Similar considerations apply to all other partonic contributions. Whenever gluons are
involved, functions analogous to the transversity, Boer-Mulders, Collins functions for
quarks, but describing linearly polarized gluons appear, see Ref.˜\ \refcite{forma}.
\begin{figure}[ht]
 \centerline{ \begin{turn}{-90}\epsfxsize=5cm\epsfbox{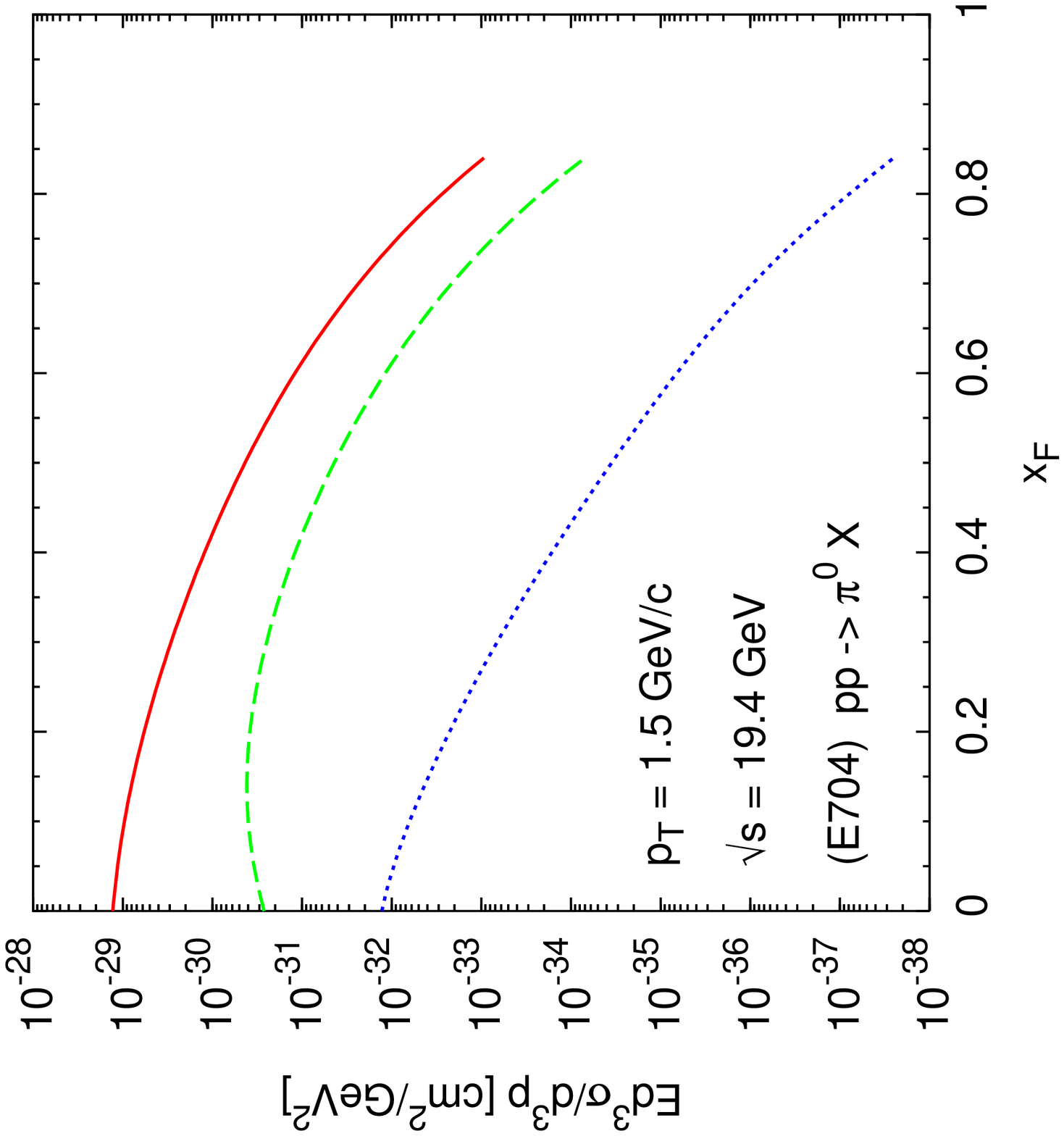}\end{turn}
 \hspace{2pt}
 \begin{turn}{-90}\epsfxsize=5cm\epsfbox{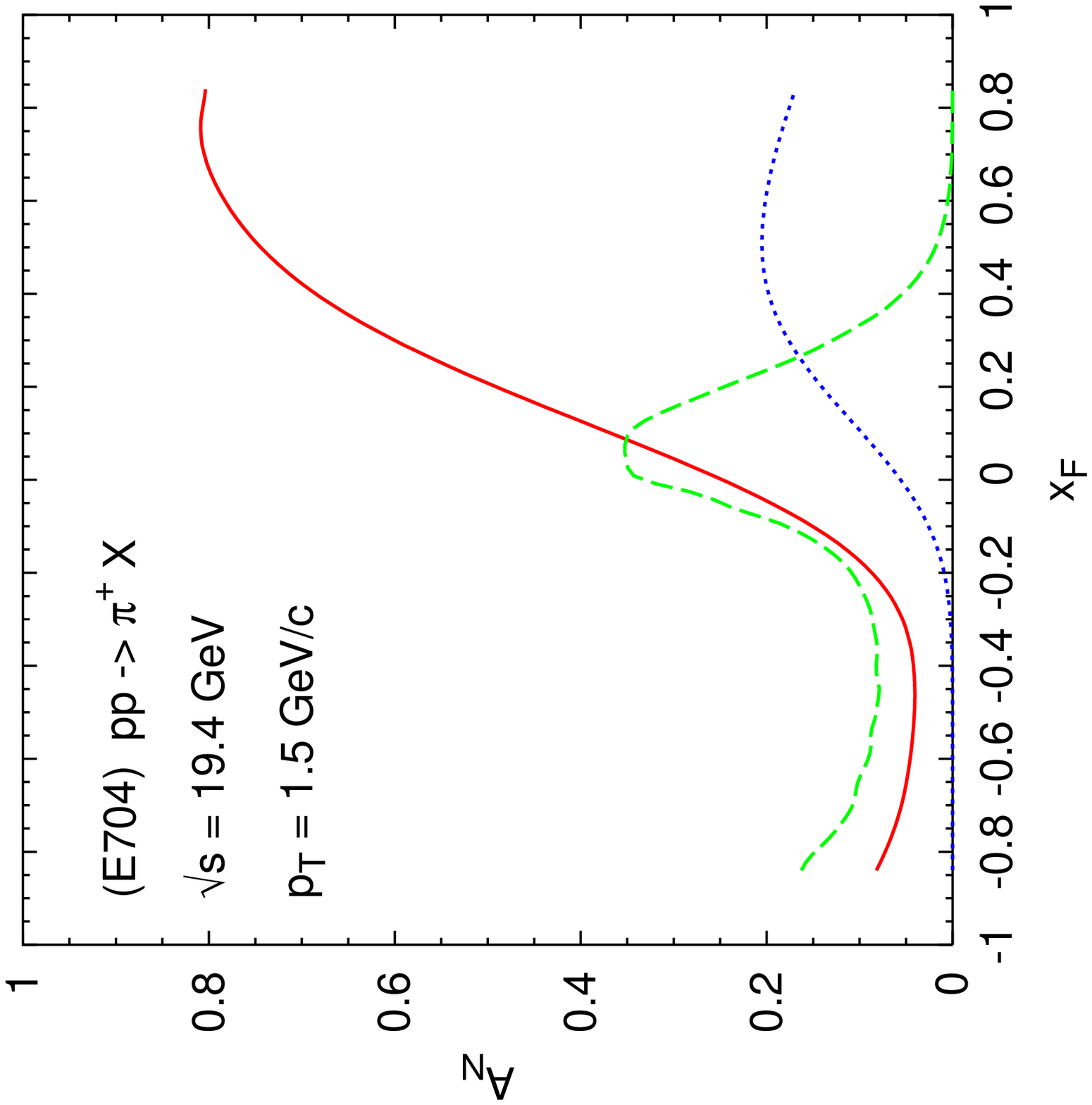}\end{turn}}
 \caption{LEFT: (maximized) contributions to the unp. cross section for $pp\to\pi^0+X$
 process and E704 kinematics; solid line: usual contribution; dashed line:
 Boer-Mulders$\otimes$Collins; dotted line: Boer-Mulders$\otimes$Boer-Mulders;
 RIGHT: (maximized) contributions to $A_N(p^\uparrow p\to\pi^+ +X)$ for E704 kinematics;
 solid line: quark Sivers contribution; dashed line: gluon Sivers; dotted line:
 transversity$\otimes$Collins; all other contributions are much smaller.
  \label{fig:E704}}
\end{figure}
 Let us now present some numerical results on unpolarized cross sections and SSA
for the $pp\to\pi+X$ process. Our aim here is basically that of showing the (maximized)
possible contributions of all terms appearing in Eqs. (\ref{num-asym-qq}),
(\ref{den-asym-qq}), which involve several unknown or poorly known functions. To this end we
saturate known positivity bounds for the Collins and Sivers functions, replacing all other
$k_\perp$ dependent polarized PDF with the corresponding unpolarized ones (keeping trace of
azimuthal phases); we sum all possible contributions with the same sign; for all PDF we
assume an $x$ and flavour independent gaussian shape vs. $k_\perp$, taking $\langle k_\perp
\rangle=0.8$ GeV/$c$, while for FF $\langle k_{\perp C}(z) \rangle$ is taken as in Ref.
\refcite{damu}; for the unpolarized, $\bfk_\perp$-integrated PDF and FF, we take respectively
the MRST01\cite{pdf} and the KKP\cite{ff} sets. See Ref.˜\ \refcite{damu} for all other
details on numerical calculations. In Fig.˜\ \ref{fig:E704} (left) we show the maximized
contributions to the unpolarized differential cross section for the $pp\to\pi^0+X$ process
for the kinematical regime of the E704 data on SSA. The usual contribution clearly dominates,
the other two being suppressed by the azimuthal phases. In Fig.˜ \ref{fig:E704} (right) we
show the contributions to the SSA, $A_N(p^\uparrow p\to\pi^+ +X)$, in the same kinematical
regime (including the negative $x_F$ region). Full $\bfk_\perp$ treatment and azimuthal
phases considerably suppress the Collins effect;\cite{abdlm} the same is not true for Sivers
contribution.
 In Fig.˜ \ref{fig:ssa} (left) we plot  $A_N(p^\uparrow p\to\pi^0 +X)$ for the kinematical regime
of the STAR experiment at RHIC. In Fig.˜ \ref{fig:ssa} (right) we show $A_N(p^\uparrow
\bar{p}\to\pi^+ +X)$, in the kinematical regime of the proposed PAX experiment at GSI. These
last results are particularly interesting for the gluon Sivers function.
\begin{figure}[ht]
 \centerline{ \begin{turn}{-90}\epsfxsize=5cm\epsfbox{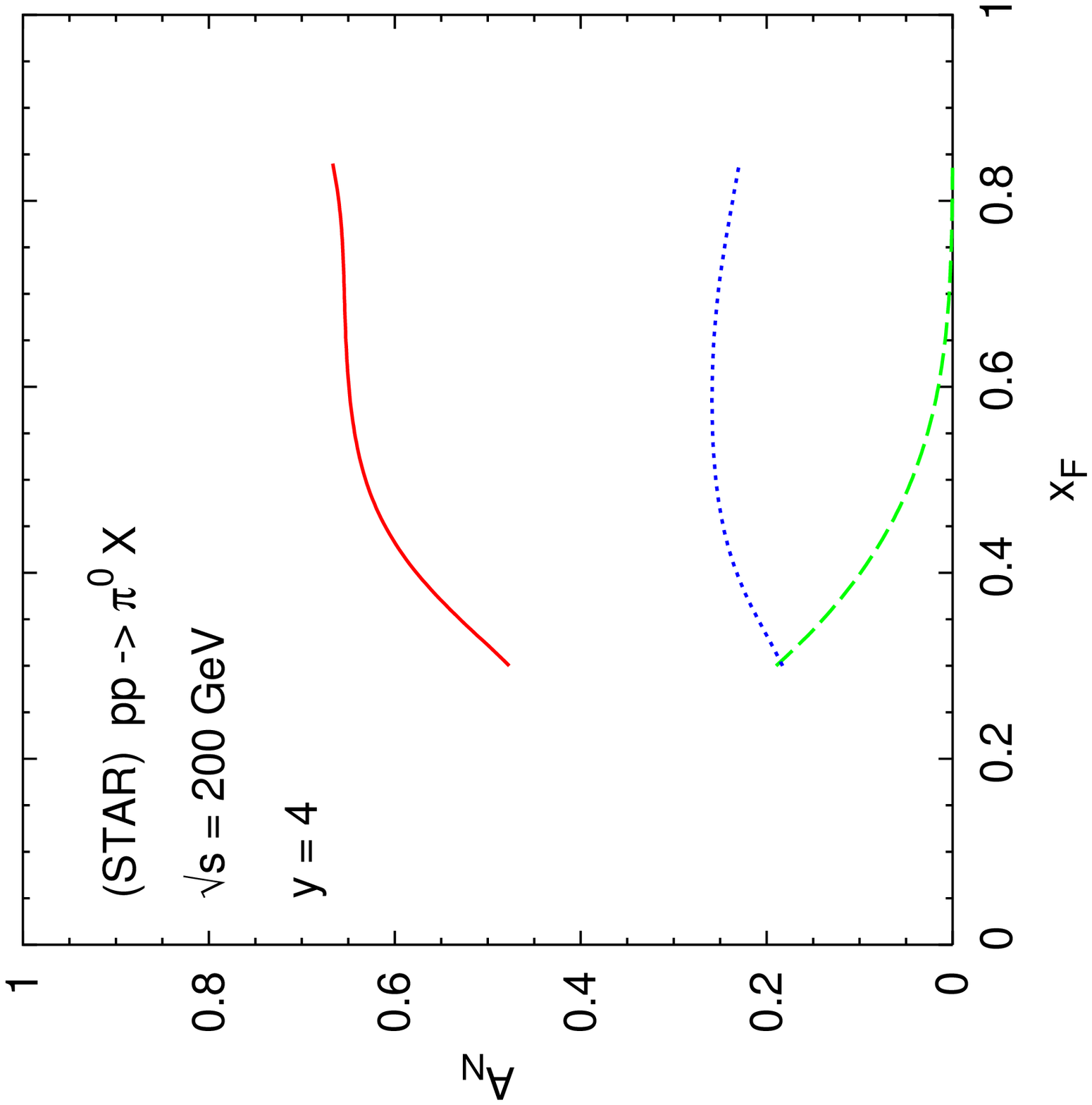}\end{turn}
 \hspace{2pt}
 \begin{turn}{-90}\epsfxsize=5cm\epsfbox{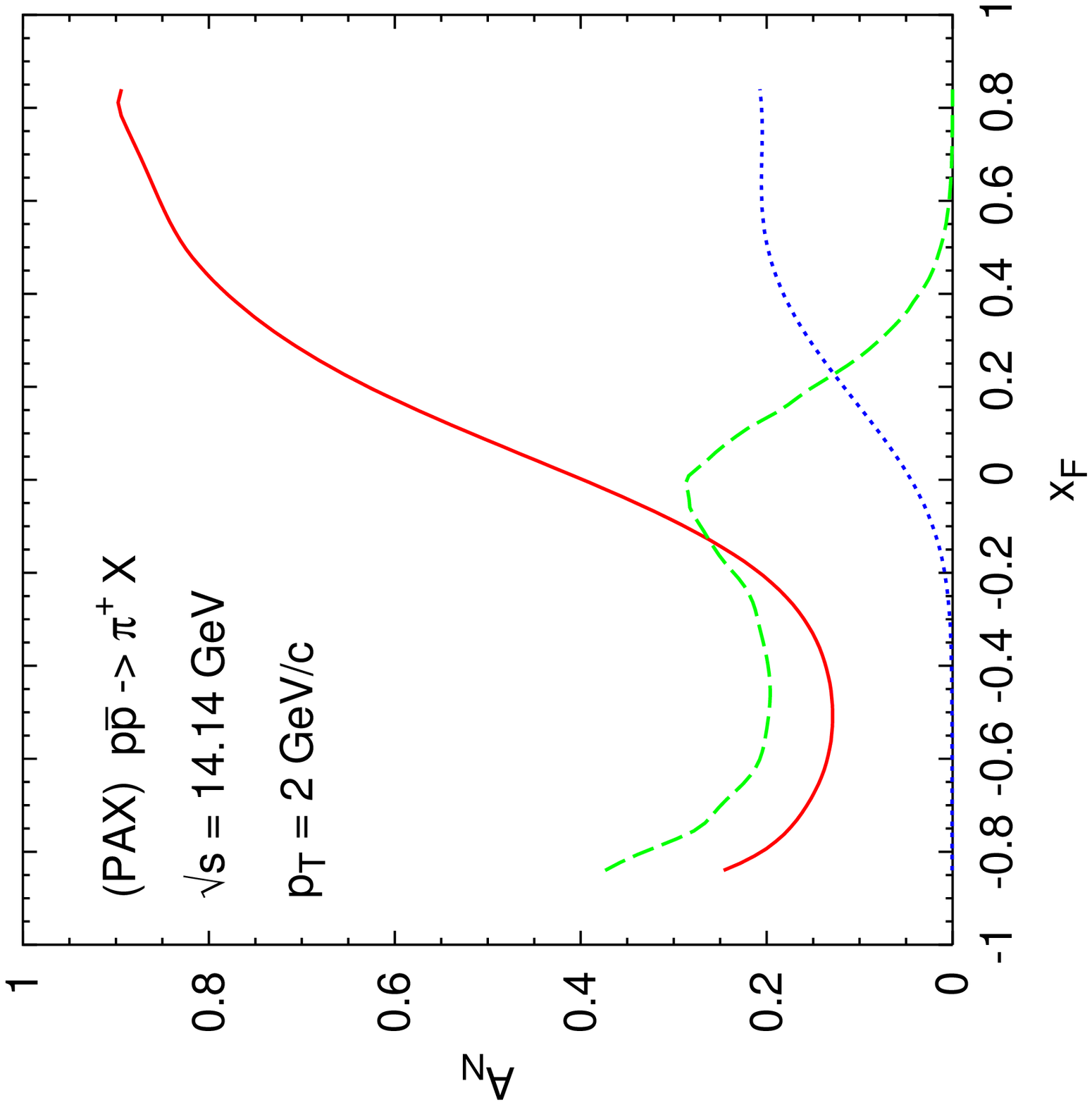}\end{turn}}
 \caption{(Maximized) contributions to: (left) $A_N(p^\uparrow p\to\pi^0 +X)$ for STAR kinematics
 (at $x_F<0$ all contributions are vanishingly small); (right)
 $A_N(p^\uparrow \bar{p}\to\pi^+ +X)$ for PAX kinematics; lines are as in Fig˜ \ref{fig:E704}.
 \label{fig:ssa}}
\end{figure}
 Many other applications of our formalism are possible and some of them are
under investigation, like the study of the double longitudinal asymmetry, $A_{LL}$, for pion
production; the transverse $\Lambda$ polarization in unpolarized hadronic reactions; the
study of the Collins effect in polarized SIDIS; the Drell-Yan process; inclusive particle
production in pion-proton collisions and so on. Hopefully, this thorough phenomenological
analysis of present and forthcoming experimental results on spin asymmetries will help in
clarifying the role of spin effects in high energy hadronic reactions.

\end{document}